\documentclass[aps,twocolumn]{revtex4-1}
\usepackage{graphicx}
\usepackage[justification=justified,width=\linewidth]{caption}
\usepackage{subcaption}
\usepackage{amsmath}
\usepackage{amsfonts}
\usepackage{amsthm}
\usepackage{amssymb}
\usepackage{amsbsy}
\usepackage{wasysym}
\usepackage{bm}
\usepackage{mathrsfs}
\usepackage{color}
\usepackage{times}
\usepackage[resetlabels]{multibib}
\begin{document}

	\title{Viscous Growth Law in Bubble Coarsening: A Molecular Dynamics Perspective}
	\author {Parameshwaran A and Bhaskar Sen Gupta}
	\email{bhaskar.sengupta@vit.ac.in}
	\affiliation{Department of Physics, School of Advanced Sciences, Vellore Institute of Technology, Vellore, Tamil Nadu - 632014, India}

\begin{abstract}
We investigate the kinetics of bubble coarsening in a single-component Lennard-Jones fluid using large-scale molecular dynamics simulations. A homogeneous high-temperature system is quenched below the critical temperature to induce the nucleation and growth of vapor bubbles within a dense liquid matrix. The structural evolution is characterized by two point correlation functions and the static structure factor, both of which exhibit dynamic scaling and sharp interfaces consistent with Porod’s law. The time-dependent characteristic length scale, extracted from the correlation function, shows a robust power law growth $\ell(t) \sim t^{\alpha}$. Finite size scaling analysis across different system sizes yields $\alpha \approx 1.0$, establishing that coarsening is dominated by viscous hydrodynamic interactions rather than classical diffusion-limited Ostwald ripening predicted by the Lifshitz-Slyozov-Wagner theory. These results provide atomistic evidence for fluid flow controlled coarsening in vapor-liquid systems and emphasize the need to go beyond diffusion-based theories to describe bubble dynamics in dense fluids.
\end{abstract}

\maketitle

\section{Introduction} 
Bubble coarsening is a fundamental non-equilibrium process that governs the late stage evolution of vapor–liquid systems and directly impacts a wide range of natural and technological phenomena. From boiling and cavitation in engineering applications \cite{zhang2017fundamentals,holmes2020experimental} to foam stability \cite{weaire1999physics,cantat2013foams,langevin2020emulsions}, volcanic degassing \cite{sparks1978dynamics}, and the processing of polymeric and metallic materials, the dynamics of bubbles strongly influence transport, mechanical properties, and energy dissipation \cite{haas2021review,miele2024bubble}. Despite its ubiquity, a comprehensive understanding of bubble coarsening remains challenging because the underlying kinetics depend sensitively on the interplay between interfacial tension, mass transfer, and hydrodynamic flow across widely separated spatial and temporal scales.

When a single-component fluid is quenched from a homogeneous high temperature state to the two-phase region below its critical temperature, it undergoes vapor-liquid phase separation driven by the amplification of spontaneous density fluctuations \cite{onuki2002phase,puri2009kinetics,bray2002theory,davis2025kinetics,parameshwaran2025kinetics,gupta2025crossover}. In this regime, the minority vapor phase typically appears as dispersed bubbles within the majority liquid phase. As the system evolves toward equilibrium, these bubbles undergo coarsening.

Although the classical Lifshitz-Slyozov-Wagner (LSW) theory~\cite{lifshitz1961kinetics,wagner1961theory} has been successfully applied to solid precipitates and droplets, its applicability to bubble coarsening in liquids remains uncertain due to the presence of strong hydrodynamic interactions. Unlike precipitates and droplets, bubbles interact via the surrounding fluid, where momentum transfer, convective flow, and pressure gradients significantly influence their motion and growth. In particular, hydrodynamic interactions introduce additional complexities that deviate from classical diffusion-driven coarsening, necessitating alternative theoretical models that explicitly account for fluid-mediated effects.

In the diffusion-dominated early time regime, the characteristic domain size \( \ell(t) \), representing the average bubble size, evolves due to the chemical potential gradient, expressed as  
\begin{equation}
\frac{d\ell(t)}{dt} \propto |\nabla \mu| \propto \frac{\gamma}{\ell(t)^2},
\end{equation}  
where \( \gamma \) denotes the interfacial tension that drives the transfer of mass between domains. Due to the higher chemical potential associated with smaller domains, the mass is redistributed from smaller to larger structures, minimizing the system’s free energy. Integration of this relation yields the well-known Lifshitz-Slyozov-Wagner (LSW) growth law $\ell(t) \sim t^{1/3}$, 
which describes diffusion-limited coarsening in phase-separating systems. However, in the intermediate and late time, when hydrodynamic interactions become significant, growth deviates from classical diffusion-limited kinetics~\cite{siggia1979late,hohenberg1977theory}. In this regime, the competition between the surface energy density \( \gamma/\ell(t) \) and the viscous stress \( 6\pi \eta v_{\ell} / \ell(t) \), where \( v_{\ell} \) is the interface velocity and \( \eta \) is the shear viscosity, dictates domain evolution. This balance yields  
\begin{equation}
\frac{d\ell(t)}{dt} \propto v_{\ell} \propto \frac{\gamma}{\eta},
\end{equation}  
which upon integration leads to a linear growth law $\ell(t) \sim t$.
This corresponds to the viscous hydrodynamic regime, which applies when the Reynolds number is low. However, as the domain size exceeds the inertial length scale, defined as \( \ell_{\text{in}} = \eta^2/(\rho \gamma) \), where \( \rho \) is the fluid density, inertial effects become dominant. In this regime, the governing balance changes from viscous stress to kinetic energy density \( \rho v_{\ell}^2 \), modifying the coarsening law to be  
\begin{equation}
\frac{d\ell(t)}{dt} \propto \frac{1}{\ell(t)^{1/2}}.
\end{equation}  
Solving this equation gives the power law growth $\ell(t) \sim t^{2/3}$,
which defines the inertial hydrodynamic regime (\( \alpha = 2/3 \)). This transition from viscous to inertial growth reflects the increasing role of momentum transport in phase-separation kinetics at larger length scales. 

Despite the significance of bubble coarsening in various industrial and natural processes, a detailed numerical investigation of hydrodynamic effects on bubble growth remains largely unexplored. In this study, we carry out molecular dynamics simulations to investigate bubble coarsening in a one-component fluid, where the complete hydrodynamic effect is taken into consideration. The main aim is to investigate the dynamics in which the vapor phase forms discrete bubbles and quantify the evolution of the bubble size, the mean bubble radius, and the associated growth exponent in the presence of hydrodynamics.

\section{Model and method} 
For the present study, we consider a single-component classical $N$ particle cubic system of volume $V$ in three dimensions, interacting via the standard Lennard-Jones (LJ) potential given as
\begin{equation}
	U(r_{ij})=4\epsilon\left[\left(\frac{\sigma}{ r_{ij}}\right)^{12}- \left(\frac{\sigma}{r_{ij}}\right)^6\right] \label{eq1} .
\end{equation}
Here, $\epsilon$ denotes the interaction strength, $\sigma$ is the particle diameter, and $r_{ij}=|r_i - r_j|$ is the scalar distance between the pair of particles $i$ and $j$. The system described above has a bulk critical temperature $T_c = 0.94\epsilon/k_B$ and a critical number density $\rho_c = 0.32$ for the vapor-liquid transition~\cite{roy2012nucleation}, $k_B$ being the Boltzmann constant. For convenience, the mass of each particle $m$, $\sigma$, $\epsilon$, $k_B$ are set to unity. The length is measured in units of $\sigma$. To improve computation speed, the LJ potential is truncated at $r_c=2.5\sigma$ and shifted to zero at $r_c$. 

To study coarsening dynamics, we resort to molecular
dynamics (MD) simulation in the canonical ensemble~\cite{AllenTildesley89}. To keep the temperature of the system constant, Nosé-Hoover thermostat is used that controls
the temperature, and at the same time, preserves the hydrodynamics of the system~\cite{FrenkelSmit02}. We consider three different cubic simulation boxes of size $L=64,96,128$ with periodic boundary conditions in all directions. The overall number density of our system $\rho =N/V= 0.7$ is chosen close to the liquid branch of the co-existence curve. All MD runs are carried out using the velocity Verlet algorithm~\cite{verlet} with time step $\Delta t = 0.005$, where the time is measured in units of LJ time $(m\sigma^2/\epsilon)^{1/2}$.

We begin our simulation by equilibrating the system at a high temperature of $T_i = 6.0\epsilon/k_B$, to prepare an initial homogeneous configuration. The evolution dynamics of the bubble-liquid phase separation is studied after quenching the system to $T_f= 0.6\epsilon/k_B (< T_c)$. The ensemble average of all statistical quantities is obtained from 20 independent runs at $T_f= 0.6\epsilon/k_B$ starting from completely different initial configurations. 

\section{Results}
At the outset, it is worth mentioning that the kinetics of bubble coarsening with hydrodynamic interactions has remained poorly explored, particularly in the context of phase separation in liquid-vapor systems. Understanding the interplay between fluid flow and coarsening dynamics is crucial, as hydrodynamics can significantly influence bubble growth, coalescence, and eventual domain evolution.  

\begin{figure}[h]
	\centering
	{\includegraphics[width=0.45\textwidth]{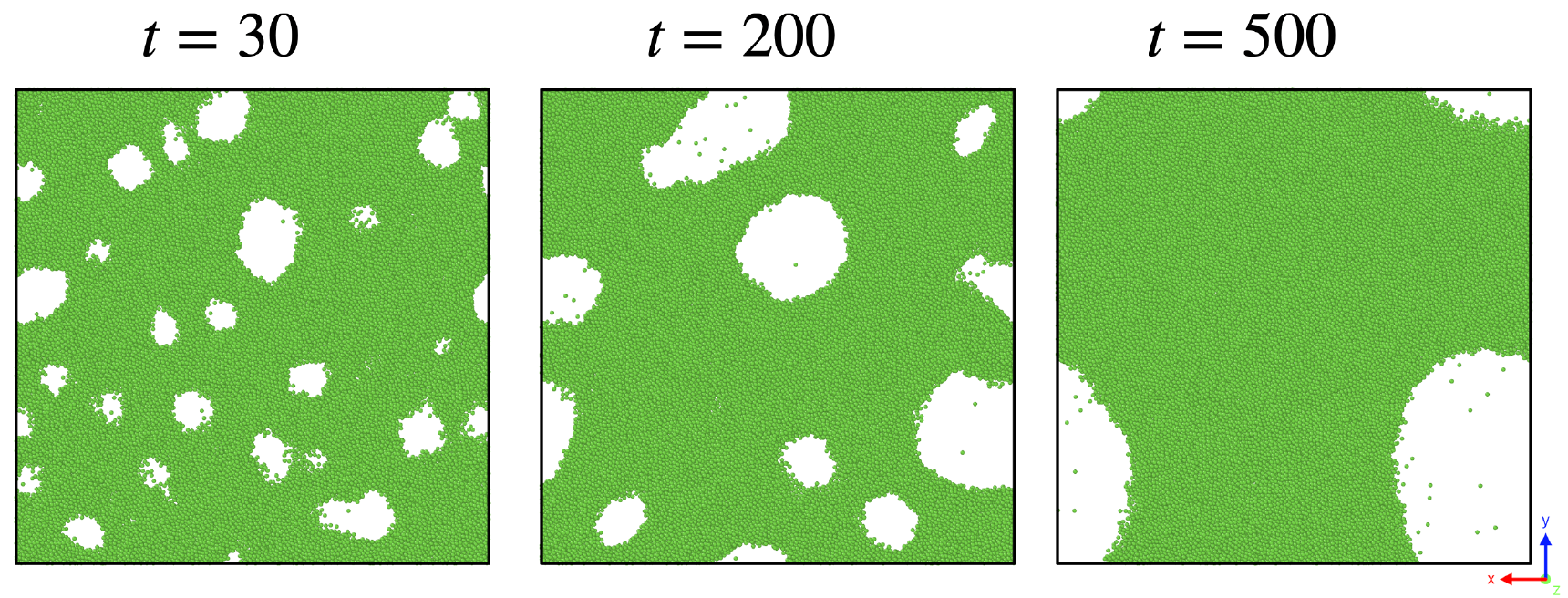}} \\
	\caption{ Time evolution of bubbles (white regions) are shown. For better visualization, two dimensional cross section of the system is displayed. The green regions represent the liquid phase.}
	\label{fig:fig1}
\end{figure}

In Fig.~\ref{fig:fig1}, we present a series of 2D cross-sectional snapshots capturing the temporal evolution of vapor bubbles following the quench of a homogeneously initialized system to a temperature of $T = 0.6\epsilon/k_B$. At this temperature, the system undergoes a phase transition, where regions of lower density emerge within the metastable liquid phase, eventually forming distinct vapor domains. Because the overall density is close to the liquid branch of the co-existence curve, the phase separation mechanism proceeds predominantly through nucleation and subsequent growth of nearly spherical vapor bubbles, rather than via spinodal decomposition.

As time progresses, the bubbles undergo Ostwald ripening, coalescing, and increasing in size, while reducing their overall number. Hydrodynamic interactions play a key role in this process, as the surrounding liquid medium facilitates momentum transfer, affecting bubble motion and growth kinetics. Additionally, surface tension-driven coarsening mechanisms ensure that smaller bubbles dissolve, feeding mass into larger ones, thereby minimizing the total interfacial energy. The competition between thermodynamic forces and hydrodynamic effects governs the coarsening dynamics, ultimately shaping the final morphology of the phase-separated system.  

For a translationally invariant system, the traditional probe to characterize configurational morphologies is the spherically averaged two-point equal time correlation function~\cite{bray2002theory,davis2025phase,davis2023surface,bhattacharyya2022}
\begin{equation}
C(r,t)= \langle \psi(0,t)\psi(\vec{r},t) \rangle  -\langle \psi(0,t)\rangle \langle \psi(\vec{r},t)\rangle
\end{equation}
where $\psi(\vec{r},t)$ is the order parameter. This is calculated by partitioning the entire system into cubic boxes of size $(2\sigma)^3 $. Subsequently, the order parameter associated with each box is assigned the value $\psi(\vec{r},t ) = +1$ if the local number density exceeds $\rho_c$ and
$\psi(\vec{r},t )= -1$ otherwise.
For patterns that are statistically self-similar, $C(r,t)$ exhibit the scaling property as 
\begin{equation}
C(r,t)\equiv \tilde{C}(r/\ell(t)),
\end{equation}
where $\tilde{C}$ is a master scaling function independent of time~\cite{bray2002theory,puri2009kinetics}. The observed scaling law facilitates the definition of the relevant time-dependent characteristic length scale, denoted as $\ell(t)$, from the decay of $C(r,t)$. In the rest of the paper we use the first 0.1 crossing of $C(r,t)$ as a measure of $\ell(t)$.
\begin{figure}[h]
	\centering
	{\includegraphics[width=0.35\textwidth]{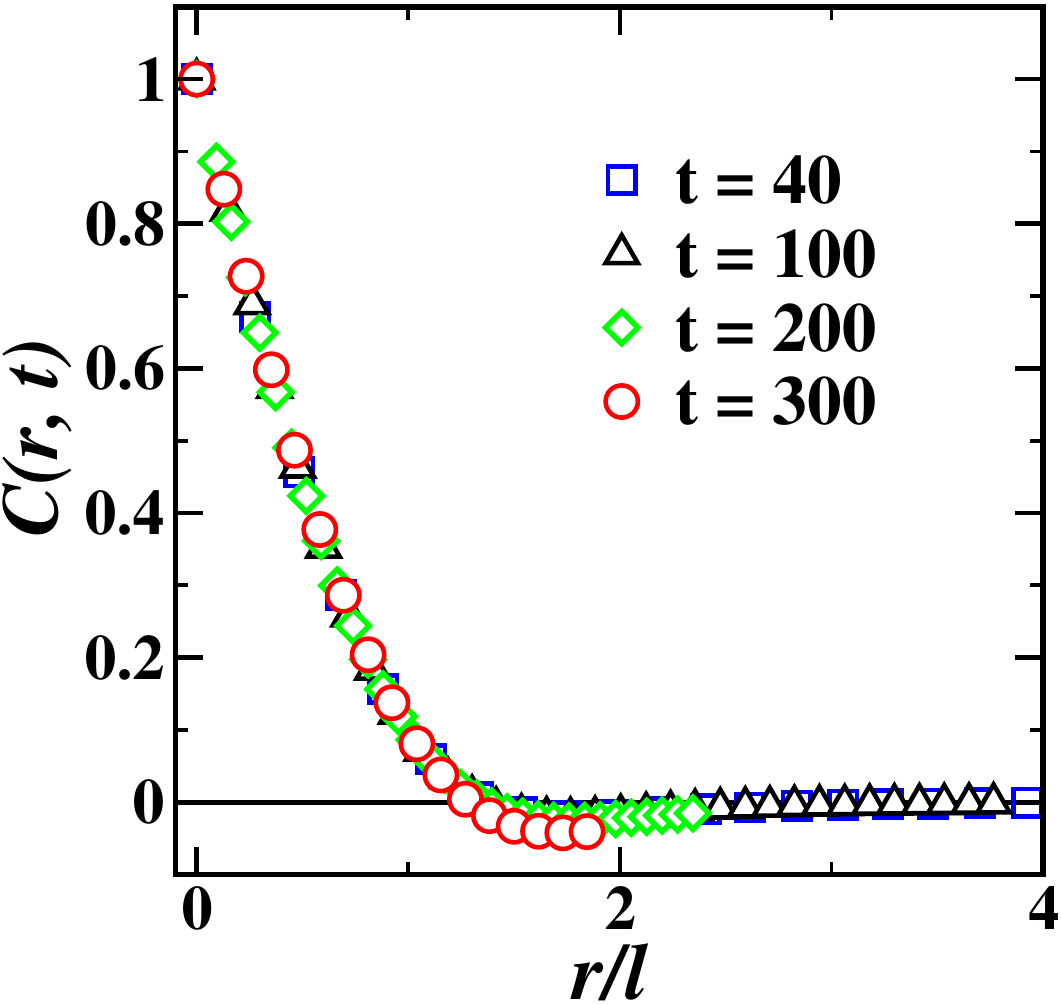}} 
	\caption{ Scaling plot of correlation $C(r,t)$ as the function of $r/\ell$ for different times.}
	\label{fig:fig2}
\end{figure}
In Fig.~\ref{fig:fig2}, we depict the scaling behavior of the correlation function $C(r,t)$ plotted against $r/\ell$. The results show a satisfactory collapse of the data across different time points, indicating that the system demonstrates consistent scaling behavior over time. The clear alignment of data across different time points highlights the self-similar behavior of pattern formation during phase separation. This suggests that the underlying physical mechanisms governing phase separation remain invariant with time, reinforcing the universality of the scaling properties observed in liquid-vapor phase separation.

To gain a deeper understanding of the morphology of domain boundaries, we compute the structure factor, which is essentially the Fourier transform of the correlation function $C(r,t)$. This transformation allows us to analyze spatial fluctuations in the phase-separated domains. The structure factor is defined as~\cite{bray2002theory}  
\begin{equation}
S(k,t) = \langle |\psi(k,t)|^2 \rangle = \int C(r,t)e^{ik \cdot r} dr,
\end{equation}  
where $k$ is the wave vector.  
For self-similar patterns, the structure factor follows a scaling form
\begin{equation}
S(k,t) \equiv \ell^d \tilde{S}(k\ell(t)),
\end{equation}  
where $\tilde{S}$ is a time-independent scaling function and $\ell(t)$ is the characteristic length scale that grows with time. This form of scaling implies that the statistical properties of the system remain unchanged when rescaled by the evolving domain size, indicating self-similar coarsening dynamics~\cite{bray2002theory,puri2009kinetics}.  

\begin{figure}[h]
	\centering
	{\includegraphics[width=0.35\textwidth]{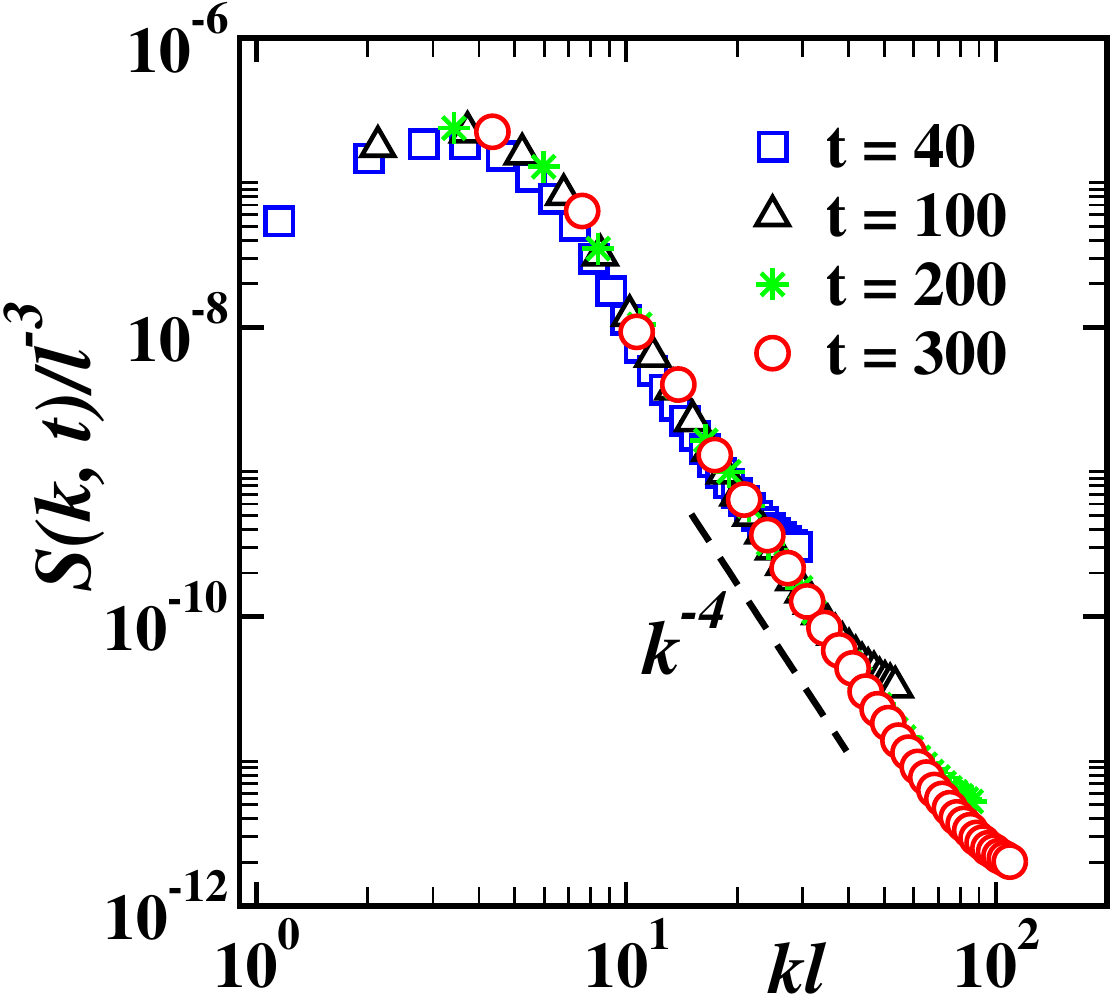}} \\
	\caption{ The scaled structure factor $S(k,t)\ell^{-3}$ vs $k\ell$ is shown for different times. The straight line is a guide line for the Porod law (see text).}
	\label{fig:fig3}
\end{figure}

In Fig.~\ref{fig:fig3}, we present a log-log plot of $S(k,t) \ell^{-3}$ versus $k\ell(t)$, incorporating data from different time steps. The data collapse observed in the figure is pretty good, supporting the validity of the scaling hypothesis. This confirms that the evolution of the structure factor follows a universal behavior.
At large $k$, the structure factor exhibits a power-law decay, following the relation  
\begin{equation}
S(k,t) \sim k^{-4}.
\end{equation}  
This decay is consistent with Porod’s law $S(k,t) \sim k^{-(d+1)}$ in $d$ dimension~\cite{bray2002theory}, that describes scattering from systems with sharp interfaces. The presence of this power-law behavior indicates that the boundaries between coexisting phases remain well defined, with sharp interfaces~\cite{bhattacharyya2021effect,bhattacharyya2024kinetics} that separate the liquid and bubble phases. The agreement with Porod’s law suggests that the phase separation mechanism is dominated by interface-driven coarsening rather than bulk diffusion.  

\begin{figure}[h]
	\centering
	{\includegraphics[width=0.35\textwidth]{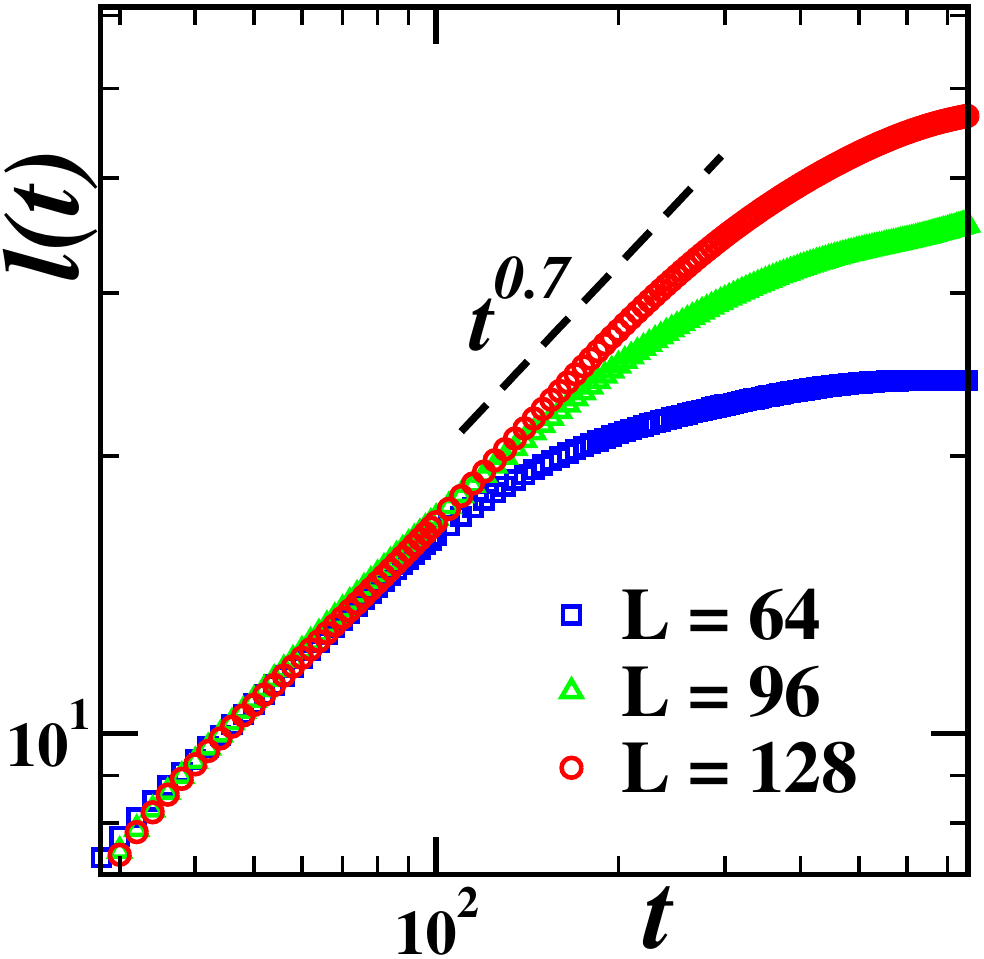}} \\
	\caption{ The time evolution of the average bubble size $\ell(t)$ is shown for three different system sizes. The dashed line represents the guideline for the slope.}
	\label{fig:fig4}
\end{figure}

The evolution of morphology during the process is best characterized by tracking the average domain size, denoted as $\ell(t)$, which serves as a key metric for understanding the system's growth dynamics. In Fig.~\ref{fig:fig4}, we present the time-dependent characteristic length scale $\ell(t)$ as a function of time $t$ for several different system sizes, plotted on a log-log scale. The data exhibit a clear power-law behavior. The presence of this power law suggests that domain coarsening follows a universal dynamical law~\cite{{bray2002theory,puri2009kinetics}}, which is largely independent of the initial conditions and specific details of the system. 
\begin{equation}
\ell(t) \sim t^{\alpha}
\end{equation}
The log-log plot in Fig.~\ref{fig:fig4} presents the time-dependent evolution of the characteristic length scale $\ell(t)$ for different system sizes ($L = 64, 96, 128$). The data clearly demonstrate a power-law growth regime, characterized by $\ell(t) \sim t^{0.7}$, as indicated by the dashed guideline. 
For small times, all system sizes follow the same universal power-law behavior, indicating that early-stage growth is independent of system size. However, at later times, deviations from this scaling behavior appear, particularly for smaller system sizes. The characteristic length $\ell(t)$ saturates earlier for smaller $L$, suggesting finite-size effects that constrain further coarsening. This saturation occurs because the system size imposes an upper limit on domain growth, preventing further evolution beyond a certain threshold. 

\begin{figure}[h]
	\centering
	{\includegraphics[width=0.35\textwidth]{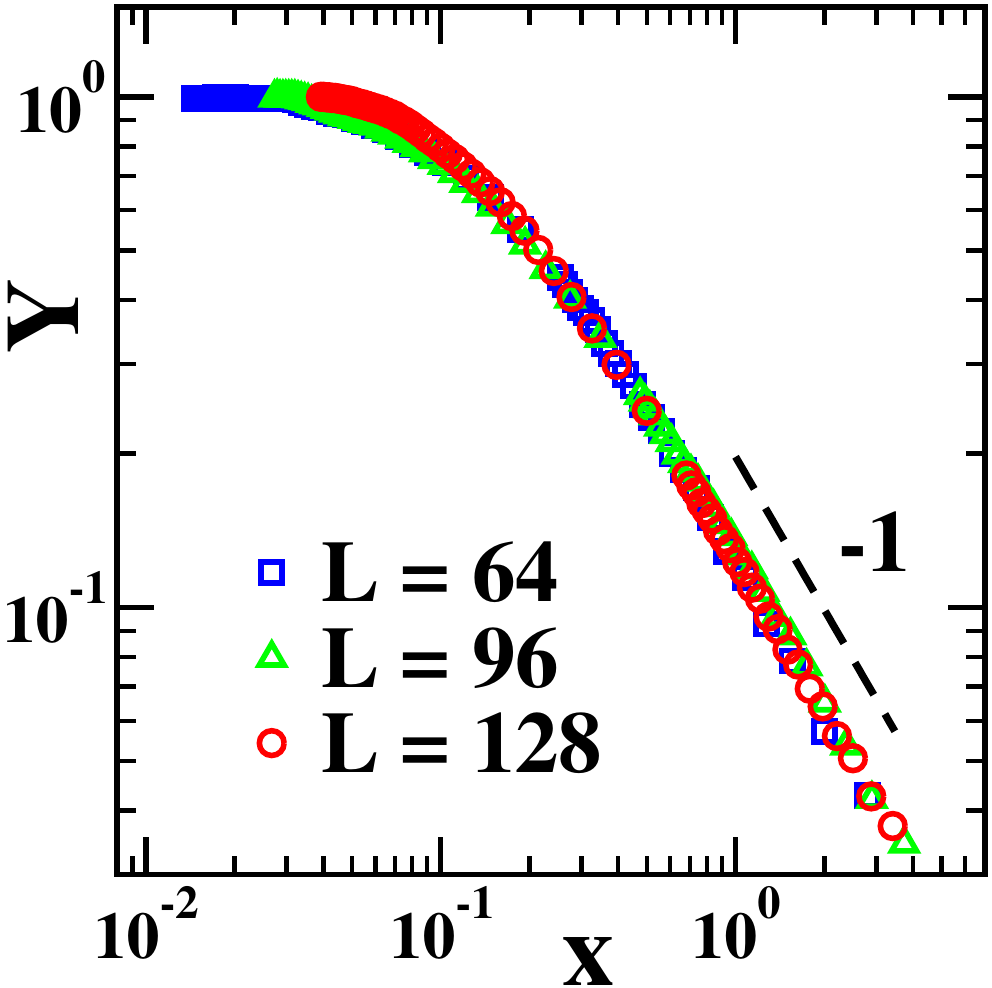}} \\
	\caption{ Finite size scaling plot for different system sizes. The dashed line represents the guideline for the slope.}
	\label{fig:fig5}
\end{figure}
Data for $L = 128$ show the most extended power-law regime before showing signs of saturation, reinforcing the importance of studying larger system sizes to capture the true asymptotic growth behavior. However, to quantitatively characterize this growth and determine the underlying scaling laws, a more systematic approach is required.

To achieve this, we employ the finite size scaling (FSS) analysis~\cite{das2006critical}, which provides a robust framework to accurately estimate the growth exponent $\alpha$. Since the system requires a time $t_0$, to reach the self-similar growth regime after a quench, the characteristic length scale can be expressed as:
\begin{equation}
\ell(t) = \ell_0 + A(t - t_0)^\alpha,
\end{equation}
where $\ell_0 = \ell(t= t_0)$ represents the characteristic length scale at time $t_0$. To analyze the late-time scaling behavior, we introduce a dimensionless scaling function $Y(X)$ as follows:
\begin{equation}
\ell(t) - \ell_0 = Y(X)(\ell_{\max} - \ell_0),
\end{equation}
where $Y(X)$ is a system size-independent function of the variable $X$:
\begin{equation}
X = \frac{(\ell_{\max} - \ell_0)^{1/\alpha}}{t - t_0}.
\end{equation}
Thus, the master curve $Y(X)$ plotted as a function of $X$, will
result in a data collapse for different system sizes, for the
appropriate choice of the parameters $\ell_0$, $t_0$ and $\alpha$. In addition,
one should obtain a power-law behavior $Y(y) \sim y^{-\alpha}$ for large X values.

 In Fig.~\ref{fig:fig5}, we plot $Y(X)$ as a function of $X$ on a log-log scale. The excellent data collapse observed across different system sizes confirms the robustness of our scaling analysis. The optimal parameters for this collapse are found to be $\ell_0 = 9.7$, $t_0 = 43$, and $\alpha = 1.0$. In the large $X$ regime, we observe a power law decay with an exponent $\alpha = 1.0$, indicating that the system exhibits a growth behavior characteristic of the viscous hydrodynamic regime. This finding strongly suggests that domain coarsening in our system is governed by hydrodynamic interactions, leading to the exponent of 1.0, which is consistent with theoretical predictions for systems dominated by fluid flow-driven dynamics.

\section{Conclusions}

In this study, we have systematically investigated the kinetics of bubble coarsening in liquid-vapor phase separation, emphasizing the role of hydrodynamic interactions in the governing domain evolution. Using molecular dynamics simulations in the NVT ensemble with a Nose-Hoover thermostat, we ensured realistic temperature regulation, thereby preserving essential hydrodynamic effects. Our analysis revealed that bubble coarsening is strongly influenced by momentum transfer, convective flows, and pressure gradients, distinguishing it from classical diffusion-limited Ostwald ripening described by the Lifshitz-Slyozov-Wagner (LSW) theory. The time evolution of the characteristic length scale, $\ell(t)$, exhibited a power law growth behavior with an exponent $\alpha  =1.0$, indicating that the system is predominantly governed by viscous hydrodynamic coarsening. Structural analysis through two-point correlation functions and structure factor scaling further confirmed the self-similar nature of bubble growth and the presence of well-defined interfaces consistent with Porod’s law. Finite-size scaling analysis validated the universality of the observed growth exponent, reinforcing the dominance of hydrodynamics in bubble coarsening. Our findings emphasize the necessity of extending beyond classical coarsening models to incorporate hydrodynamic effects, particularly for liquid-bubble phase separation in high-density systems. These insights have broad implications for understanding phase separation dynamics in foams, cavitation, boiling processes, and industrial gas-liquid systems. Future work could explore the impact of external factors such as shear flow, surfactants, and thermal gradients on bubble coarsening, further enriching our understanding of fluid-mediated phase separation.\\

\textit{Acknowledgements.}--B. Sen Gupta acknowledges Science and Engineering Research Board (SERB), Department of Science and Technology (DST), Government of India (no. CRG/2022/009343) for financial support. Parameshwaran A. acknowledges DST-SERB, India for doctoral fellowship.

\end{document}